# Surface plasmon resonance of Cu nanowires in polycarbonate template

## A. Azarian[a,*] and F. Babaei[a]

[a] *Department of Physics, University of Qom, P.O. Box 37185 - 359, Qom ,Iran*
[*] *Email: azarian @qom.ac.ir; Tel: +98 251 2853311; Fax: +98 251 2854972*

## Abstract

The Cu nanowires were electrodeposited in polycarbonate track-etched (PCT) membrane. SEM, TEM and XPS techniques were used to characterize the morphology, structure and size of nanowires as well as chemical Composition. The absorption spectrum of copper nanowires embedded in PCT was measured and calculated for different incident angles and wavelengths. Our results showed that there is a broad peak due to excitation surface plasmon at $\theta = 70°$ for wavelength $\lambda = 730$ nm. We applied the transfer matrix method and Bruggeman homogenization formalism for optical modeling. The results of absorption spectra showed that exist good agreement between the experimental and our used model. The results of this work may be useful in study of surface plasmon resonance of copper nanowires.

## 1- Introduction

In recent years, metal nanowires have found great attention because of the variety of applications such as inter Connections in nanoelectronic, probes, biological sensors, storage media and sensing devices [1-3]. Broad investigations of surface plasmon resonance have been done at the interface between a metal and an isotropic dielectric material [4, 5]. The optical properties of metallic nanostructures and interaction of light with nanometer size particles and wires have attracted a lot of attention recently due to their potential applications in nonlinear optics and plasmonics [6-9]. Those nanostructures have been intensely investigated as substrates for surface enhanced vibrational spectroscopy (SEVS), such as the surface enhanced Raman spectroscopy (SERS)[10-12], the surface enhanced infrared absorbance (SEIRA)spectroscopy[13]. Localized surface plasmon resonance is one of the major mechanisms of mentioned spectroscopies. The existence of sharp peak in optical absorbance of structure as angle of incident of light with linear polarization is due to excitation surface plasmon( SP) wave [14].

SP waves are used for high-throughput analysis of biomolecular interactions for proteomics, drug discovery, and pathway elucidation [15, 16], sensing [17], for high-speed addressing of information on computer chips [18] and long range-communications [19].

One simple and versatile approach for preparing of metal nanowires is the template method. Today, the most suitable template for metal nanowires is cylindrical narrow pores of polycarbonate track-etched (PCT) membranes [20], anodic aluminum oxide (AAO) templates [21] and nano pores mica [22]. The PCT template is an ideal structure, because it has tunable pore dimensions over a wide range of diameters and shows better wetting properties relative to AAO template. In the present work, Cu nanowires were electrodeposited in PCT template and then absorption spectrum was measured for different wavelengths and incident angles.

In this study, we report on the surface plasmon resonance of the Cu nanowires embedded in polycarbonate membrane. The fabrication method and optical modeling are outlined in Section 2 & 3, respectively. Results are presented and discussed in Section 4.

## 2- Experimental

The arrays of Cu nanowires were obtained by electrodeposition of copper inside the pores of commercial polycarbonate track-etched membrane. The polycarbonate membranes used in this work had a thickness d ≈ 4μm and pores diameter of 50 nm, respectively. The growth of nanowires was performed by electrodeposition at room temperature from sulfate bath containing $Cu^{2+}$ ions. Before electrodeposition, a gold thin film with about 50 nm thicknesses were deposited on the back side of PCT templates by the sputtering technique, serving as working electrode. The used electrolyte to electrodeposition the Cu nanowires had the following composition: 50 gr / lit $CuSO_4$. 5 $H_2O$ and 30 gr / lit $H_3BO_3$ (pH 3).

The electrodeposition were performed using an EG&G potentiostate and in conventional three electrode cell. For electrodeposition, a saturated calomel electrode (SCE) was used as reference for the applied potential. More over a Pt plate was immersed in the electrolyte solution as the counter electrode, which is placed 7.0 mm away from the surface of the working electrode (PCT membrane). The current is measured during electrodeposition at a fixed potential versus SCE. Scanning Electron Microscopy (SEM) and Transmission Electron Microscopy (TEM) were used to characterize the morphology, structure and the size of nanowires. For TEM observations, a piece of Cu nanowires embedded in PCT was immersed in $CH_2Cl_2$ solution to remove the PCT membrane completely. A drop of solution was placed on a grid and allowed to dry prior to electron microscope analysis. X-ray photoelectron spectroscopy (XPS) analysis was also carried out in an ultra-high-vacuum chamber for the dispersed Cu nanowires on the Si substrate. The residual gas pressure in the analyzing chamber was $10^{-8}$ Torr. The monochromatized Al Kα (1486.6 eV) source was operated at 80 W. Optical transmission experiments were carried out using Jusco UV-Vis spectrometer.

# 3- Optical modeling

The Kretschmann configuration is a common experimental arrangement for the excitation and detection of SP waves [4, 5]. In order to excitation surface plasmon, we used a configuration similar to that Kretschmann configuration is as follows: the region $0 \leq z \leq d_1$ is occupied by a metal of relative permittivity $\varepsilon_{met}$, the region $d_1 \leq z \leq d_2$ is described by Cu nanowires (the angle of rise of nanowire relative to substrate is $\chi$) embedded in polycarbonate membrane, whereas the regions $z \leq 0$ and $z \geq (d_1 + d_2)$ are homogeneous isotropic dielectric material of relative permittivity $\varepsilon_l = n_l^2$, as it is shown in Fig.1.

A plane wave in the half space $z \leq 0$ propagate at an angle $\theta_{inc}$ to the z- axis and at an angle $\psi_{inc}$ to the x- axis in the xy - plane. The phasors of incident, reflected and transmitted electric fields are given as [23]:

$$\begin{cases} \underline{E}_{inc}(\underline{r}) = [a_s \underline{S} + a_p \underline{P}_+] e^{i\underline{K}_0 n_l \cdot \underline{r}}, & z \leq 0 \\ \underline{E}_{ref}(\underline{r}) = [r_s \underline{S} + r_p \underline{P}_-] e^{-i\underline{K}_0 n_l \cdot \underline{r}}, & z \leq 0 \\ \underline{E}_{tr}(\underline{r}) = [t_s \underline{S} + t_p \underline{P}_+] e^{i\underline{K}_0 n_l \cdot \underline{r}} e^{i\underline{K}_0 n_l \cdot (\underline{r} - l_\Sigma \underline{u}_z)}, & z \geq (d_1 + d_2) \end{cases} \quad (1)$$

The magnetic field's phasor in any region is given as:

$$\underline{H}(\underline{r}) = (i\omega\mu_0)^{-1} \underline{\nabla} \times \underline{E}(\underline{r})$$

where $(a_s, a_p)$, $(r_s, r_p)$ and $(t_s, t_p)$ are the amplitudes of incident plane wave, and reflected and transmitted waves with S- or P- polarizations. We also have;

$$\begin{cases} \underline{r} = x\underline{u}_x + y\underline{u}_y + z\underline{u}_z \\ \underline{K}_0 = K_0(\sin\theta_{inc}\cos\psi_{inc}\,\underline{u}_x + \sin\theta_{inc}\sin\psi_{inc}\,\underline{u}_y + \cos\theta_{inc}\,\underline{u}_z) \end{cases} \quad (2)$$

where $K_0 = \omega\sqrt{\mu_0\varepsilon_0} = 2\pi/\lambda_0$ is the free space wave number, $\lambda_0$ is the free space wavelength, $\varepsilon_0 = 8.854 \times 10^{-12}\,Fm^{-1}$ and $\mu_0 = 4\pi \times 10^{-7}\,Hm^{-1}$ are the permittivity and permeability of free space (vacuum), respectively. The unit vectors for linear polarization normal and parallel to the incident plane, $\underline{S}$ and $\underline{P}$, respectively are defined as:

$$\begin{cases} \underline{S} = -\sin\psi_{inc}\underline{u}_x + \cos\psi_{inc}\underline{u}_y \\ \underline{P}_\pm = \mp(\cos\theta_{inc}\cos\psi_{inc}\underline{u}_x + \cos\theta_{inc}\sin\psi_{inc}\underline{u}_y) + \sin\theta_{inc}\underline{u}_z \end{cases} \quad (3)$$

and $\underline{u}_{x,y,z}$ are the unit vectors in Cartesian coordinates system.

The reflectance and transmittance amplitudes can be obtained, using the continuity of the tangential components of electrical and magnetic fields at interfaces and solving the algebraic matrix equation [24, 25]:

$$\begin{bmatrix} t_s \\ t_p \\ 0 \\ 0 \end{bmatrix} = [\underline{\underline{K}}]^{-1} \cdot [\underline{\underline{M}}_{Cu-PCT}] \cdot [\underline{\underline{M}}_{met}] \cdot [\underline{\underline{K}}] \cdot \begin{bmatrix} a_s \\ a_p \\ r_s \\ r_p \end{bmatrix} \quad (4)$$

The different terms and parameters of this equation are given in detail by Lakhtakia and Messeir [25]. The reflection and transmission can be calculated as:

$$\begin{cases} R_{i,j} = \left|\dfrac{r_i}{a_j}\right|^2 \\ T_{i,j} = \left|\dfrac{t_i}{a_j}\right|^2 \end{cases}, i,j = s,p \quad (5)$$

The absorbance for linear polarizations s, p is calculated as:

$$A_i = 1 - \sum_{j=s,p} R_{ji} + T_{ji} \quad , i = s,p \quad (6)$$

The absorbance for natural light is calculated as:

$$A = \frac{A_s + A_p}{2} \quad (7)$$

## 4- Results and discussion

Typical current transient curve for electrodeposition of Cu nanowires in PCT is shown in Fig.2. At initial times of deposition, the current decreases due to mass transport limitation (stage *I*). In stage *II*, the metal is growing in the pores and a slightly increasing current is observed. When the pores are filled and wires reach on the top of membrane surface, because of three dimensional depositions, the effective surface area of cathode increases and a rapidly increasing deposition current can be observed (stage *III*). In this stage the hemispherical caps originating from each nanowire (this stage is not shown in figure 2). By stopping electrodeposition before starting stage *II*, Cu nanowires fill up most of the pores.

In Fig. 3 SEM micrograph of cross section of the Cu nanowires embedded in PCT was showed at the end of stage *III*. Fig. 3 confirms that the Cu nanowires growth vertically to the surface of substrate and two nanowires which are pointed out by elliptic shape. Fig. 4 shows a typical TEM image of the Cu nanowires with the average length of about 4 μm. Diameter of the nanowire at both ends is about 50 nm, same diameter of used membrane's pores, but in the middle the diameter is about 100 nm. This is due to shape of polycarbonate template pores.

Fig. 5 shows the XPS spectrum for Cu nanowires dispersed on Si substrate. The acquired XPS spectrum is analyzed by the Gaussian-Lorantzian fitting method. A dual peak is observed at about 933 eV corresponding to Cu 2p3, due to the presence of oxidation states. Amount of copper in compound form was estimated about 19 % from the area of dual peak. The excess oxygen is related to the formation of oxidation at the surface, probably after exposure to air.

To study the optical properties of wires we measured absorption spectrum Cu nanowires embedded in PCT at different wavelengths. Figure 6 demonstrates typical absorption spectra versus incident angle for wavelength $\lambda = 730$ nm. It is clear from this figure that absorption spectra increases slightly slop versus incident angle for $\theta_{inc} \prec 65^0$, but for higher values of $\theta_{inc}$ the absorption increase rapidly. It should be mentioned that for technical problems we could not measure absorption spectra for polar angles higher than $70^0$. It is well known that this behavior related to surface plasmon resonance of nanowires [14].

In optical modeling, the relative permittivity scalars $\varepsilon_{a,b,c}$ of Cu nanowires in polycarbonate template (Cu-PCT) were calculated using the Bruggeman homogenization formalism [26], as it is shown in Fig.7. In this formalism, the structure is considered as a two component composite (copper and polycarbonate). These quantities are dependent on different parameters, namely, columnar form factor, fraction of copper ($f_{Cu}$), the wavelength of free space and the refractive index $n(\lambda_0) + ik(\lambda_0)$. In addition, each column in the structure is considered as a string of identical long ellipsoids [25]. The ellipsoids are considered to be electrically small (i.e. small in a sense that their electrical interaction can be ignored) [27]. In all calculations columnar form factors $(\frac{c}{a})_{Cu} = (\frac{c}{a})_{PCT} = 20$, $(\frac{b}{a})_{Cu} = (\frac{b}{a})_{PCT} = 1.06$ (c, a and b are semi major axis and small half-axes of ellipsoids) [28] and experimental structural parameters of Cu-PCT $\chi = 90°$, $d_{Cu-PCT} = 4\,\mu m$, $f_{CU} = 0.008$ were fixed. Setting the shape factors $(\frac{c}{a})_{Cu\,\&\,PCT} \gg 1$ and $(\frac{b}{a})_{Cu\,\&\,PCT} \gg 1$ will make each ellipsoid resemble a needle with a slight bulge in its middle part [25]. we have used the bulk experimental refractive indexes copper and polycarbonate for homogenization [29].

The calculated optical absorption of structure as a function of wavelength at different polar angles for natural light plane wave is plotted in Fig 8. The absorption spectra does not show a peak until $70^0$ as one could see in Fig.8. However, in $\theta_{inc} = 70^0$ a broad peak appears in absorption spectra.

This peak shifts to shorter wavelengths for polar angles higher than $70^0$. Fig. 9 shows typical experimental optical absorption spectra of Cu nanowires embedded in PCT at incident angle of θ=70. As it is clear, there is a peak at wavelength of 773 nm which confirms results of our theoretical model at Fig 8 with a peak at 775 nm. Although, the full wide at half maximum (FWHM) of experimental peak is only 8 nm which is smaller of theoretical peak (FWHM= 50 nm). In Fig 10 the calculated absorption spectra as a function of polar angle is depicted at wavelengths 650,700,730 and 750nm. The results showed that there is a broad peak in $\theta_{inc} \approx 70^0$ at wavelength 700 nm. As mentioned above because of technical problems we could not measure absorption spectra for $\theta_{inc} \succ 70^0$. By comparison of Figs.6 and 10 it is seen that there are qualitative agreement between experimental data and results of our applied optical modeling.

## 4- Conclusions

In summary, we have successfully fabricated Cu nanowires in polycarbonate membrane by electrodeposition method. The diameter of Cu nanowires varies between 50 nm at both ends to 100 nm in the middle of each wire. The current transient curves for electrodeposition process was measured to Control the nanowires' length. The absorption spectrum of the Cu nanowires embedded in PCT shows a broad shoulder (due to excitation surface plasmon) at θ = 70 for wavelength of 730 nm. The results of optical modeling showed that there is a peak in $\theta_{inc} \approx 70^0$ at wavelength about 700 nm and our calculations are consistent with experimental results. The results of this work may be applied to the plasmonic characterization of copper nanowires.

## Acknowledgements

This work was carried out with the support of the University of Qom.

**Figure captions**

Figure 1. Schematic of the structure of thin film for optical modeling.

Figure 2. Chronoampermetry curve of electrodeposition, potential -1.3 V.

Figure 3. Cross section of the Cu nanowires embedded in PCT at the end of stage ***III***.

Figure 4. TEM image of the Cu nanowires with average length about 4 μm.

Figure 5. Cu 2p3 peak of XPS spectrum of the Cu nanowires.

Figure 6. Absorption spectrum of the Cu nanowires embedded in PCT for different incident angles at λ= 730 nm.

Figure 7. Calculated the relative permittivity scalars $\varepsilon_a$, $\varepsilon_b$ and $\varepsilon_c$ composite medium. The Re[ ] and Im[ ] denote to real and imaginary parts the relative permittivity scalars.

Figure 8. Calculated absorbance as function of wavelength, when $\psi_{inc} = 0^0$, at different $\theta_{inc}$ for natural light plane wave. The structure is described by following parameters: $d_1 = 50 nm$, $d_2 = 4 \mu m$, $\varepsilon_l = 2.9$ (silicon oxynitride $SiO_{.4}N_{.6}$), $f_{Cu} = 0.008$ and $n_{PCT} = 1.584$.

Figure 9. Typical experimental optical absorption spectra of Cu nanowires embedded in PCT at incident angle of θ=70.

Figure 10. Calculated absorbance as function of $\theta_{inc}$ when $\psi_{inc} = 0^0$, at wavelengths 650, 700, 730 and 750 nm for natural light plane wave. All other parameters of the structure are the same as for Fig. 7.

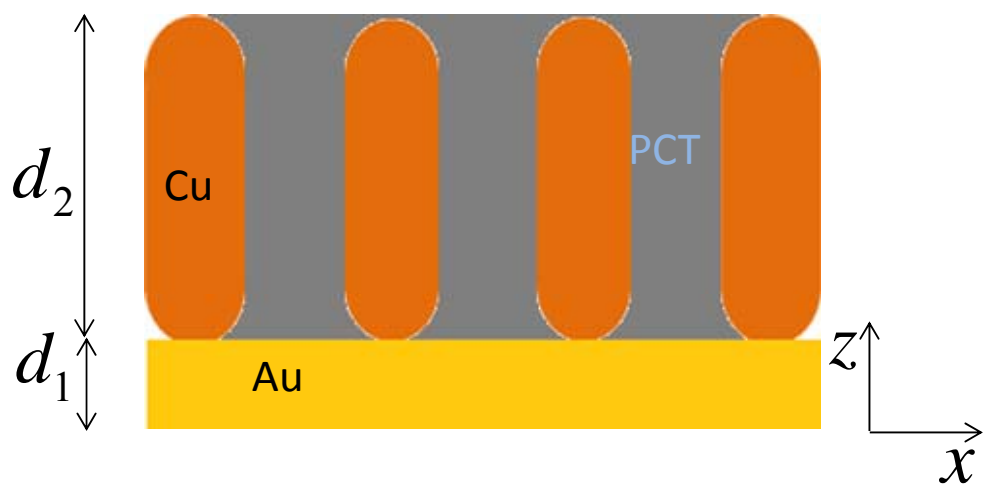

**Fig. 1; A. Azarian and F. Babaei**

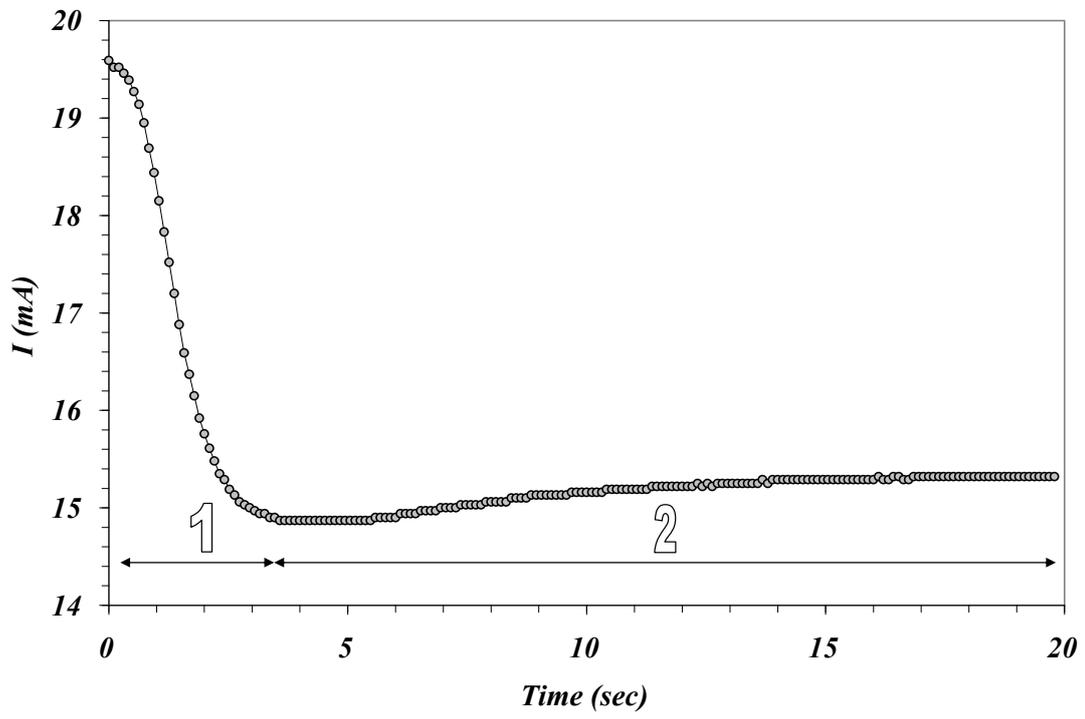

**Fig. 2; A. Azarian and F. Babaei**

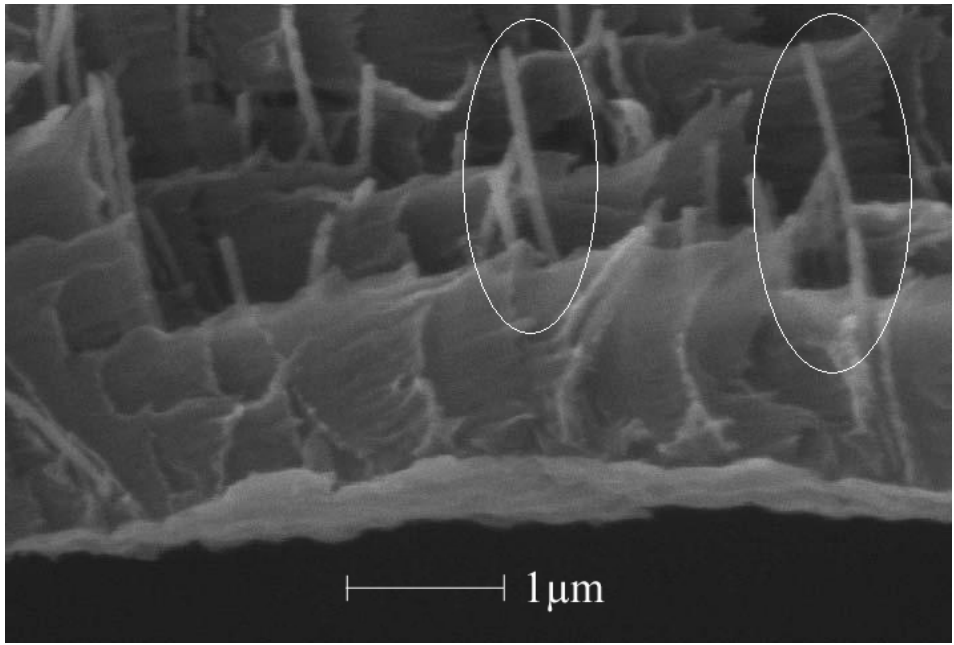

**Fig. 3; A. Azarian and F. Babaei**

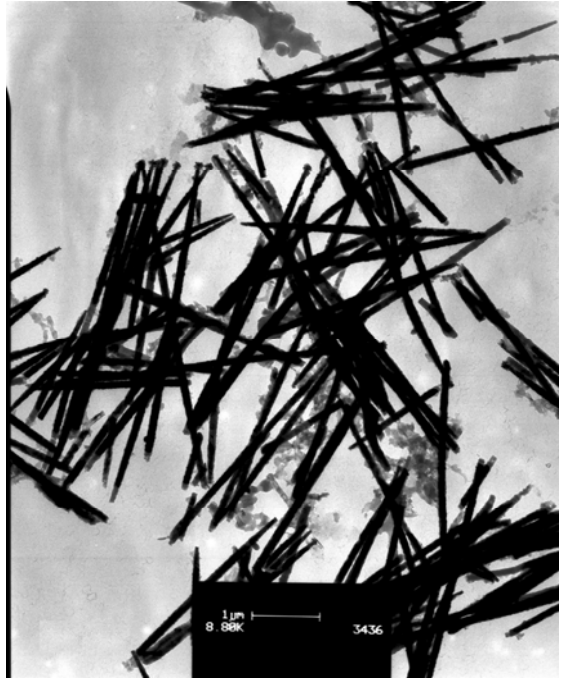

**Fig. 4; A. Azarian and F. Babaei**

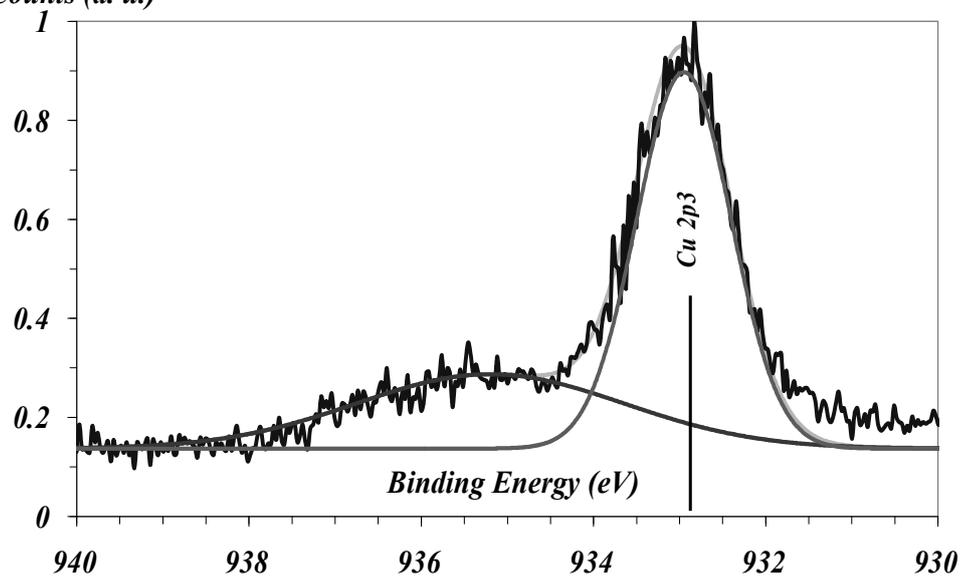

**Fig. 5; A. Azarian and F. Babaei**

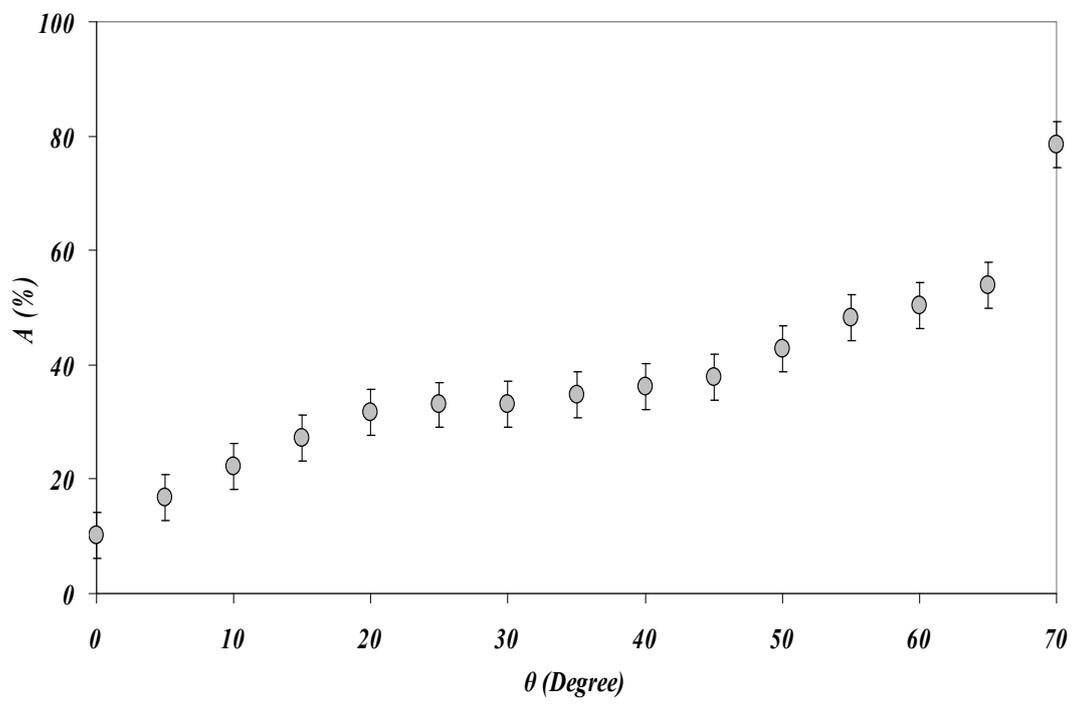

**Fig. 6; A. Azarian and F. Babaei**

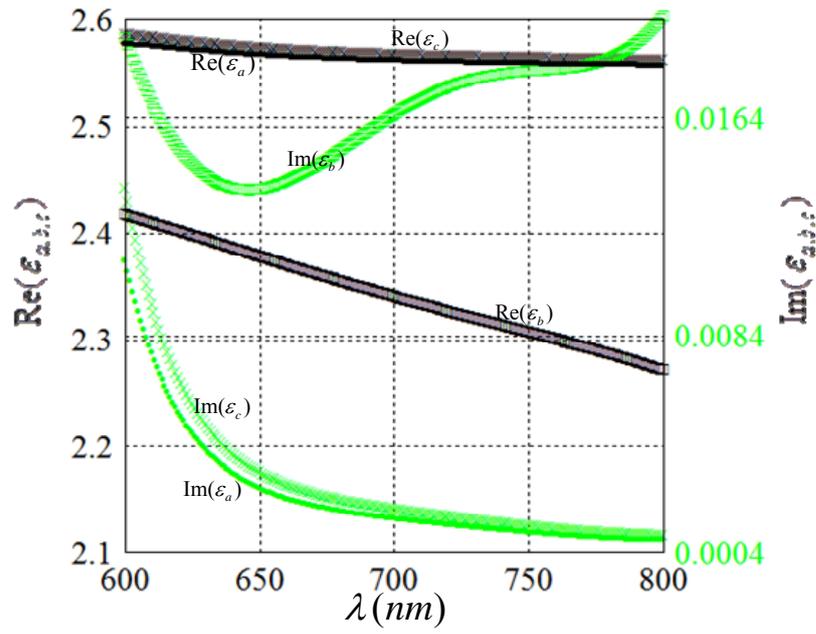

**Fig. 7; A. Azarian and F. Babaei**

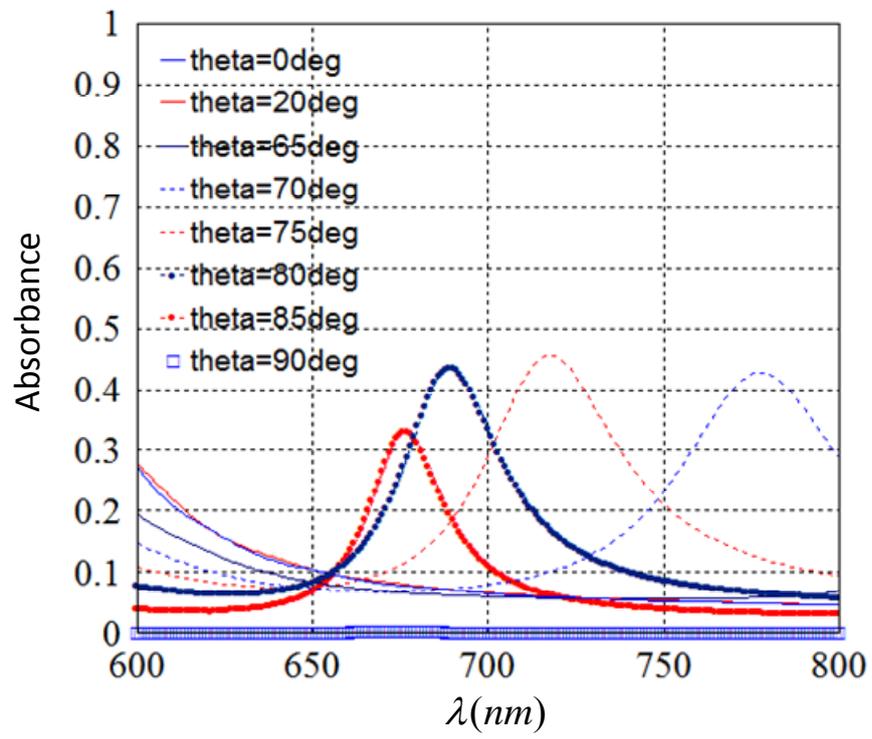

**Fig. 8; A. Azarian and F. Babaei**

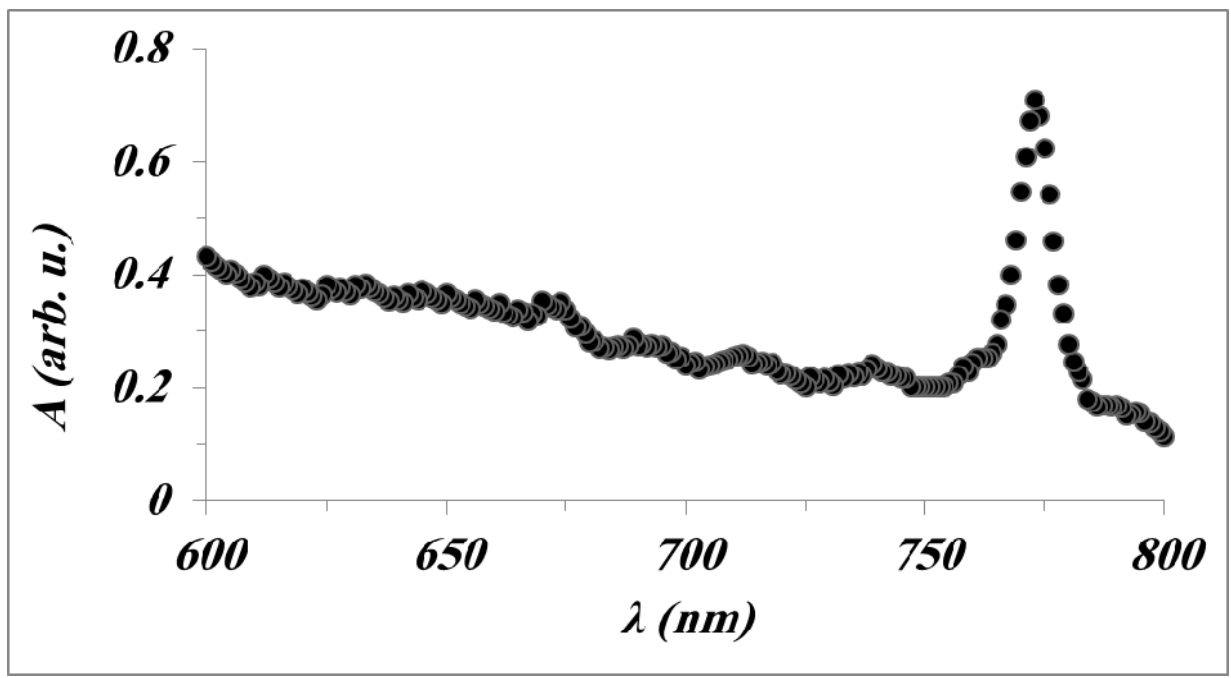

**Fig. 9; A. Azarian and F. Babaei**

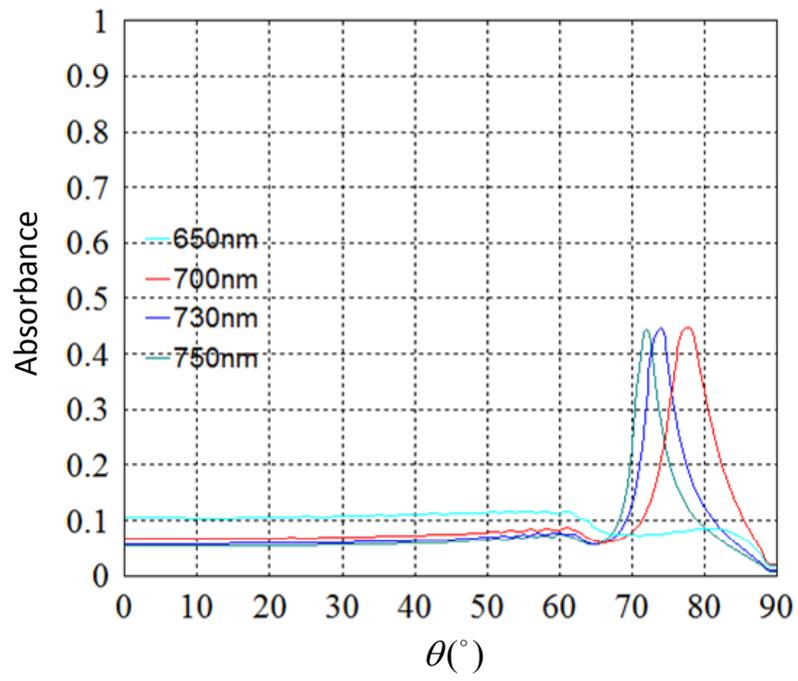

**Fig.10; A. Azarian and F. Babaei**